\documentclass[11pt,leqno,addw]{article}
\usepackage[cp1251]{inputenc}
\usepackage[russian]{babel}

\begin{document}

\begin{center}
{\bf \large{Волновая функция и распределения наблюдаемых в
релятивистской квантовой теории}}
\vspace*{1.5mm}

В.Ф.~Кротов\\
\vspace*{1.5mm}

{\em Институт проблем управления Российской академии наук,
Профсоюзная, 65, 117997, Москва, Россия}\\

(Дата: Март 16, 2009)
\end{center}
\vspace*{1mm}

{\scriptsize{Определена плотность  вероятности координат
релятивистской частицы, удовлетворяющая формальным требованиям
квантовой механики и специальной теории относительности (согласно
хрестоматийному мнению,  такая плотность отсутствует). Это~---
релятивистский инвариант, описывающий вероятность появления
частицы в пространстве-времени. Она конкретизируется  для бозона,
нейтрального и заряженного, скалярного и векторного, включая
фотон, электрона. Доказана единственность такой конструкции. Время
лишается функции динамического параметра распределения и формально
уравнивается с другими координатами. Уточняется также смысл
распределения 4-импульса. Рассматриваются условия наблюдения таких
распределений. Применительно к квантовому полю эти конструкции
трансформируются в новые характеристики распределения частиц в
пространстве-времени, дополняющие таковые для импульсных
состояний. Получены их операторы для свободного поля. В рамках
этой модели отношения неопределенностей Гейзенберга получают
теоретическое обоснование применительно к релятивистской частице;
частицы и античастицы оказываются разными состояниями одной
частицы; вакуумное состояние бозонного квантового поля  не
обладает энергией.}}
\vspace*{1.5mm}

{\flushleft{PACS  03+p; 03.65-w; 03.65 Ta}}

\section*{1. Введение}

\hspace{0.5cm} В модели квантовой системы (КС) синтезированы
детерминированная динамика волновой функции (ВФ) и статистическая
связь последней с наблюдаемыми величинами: динамическая и
статистическая части ее описания. Если первая вполне
формализована, то проблемы второй для релятивистской квантовой
частицы (РКЧ) привели к общему мнению, что $\ll$в последовательной
релятивистской квантовой механике (КМ) координаты частиц вообще не
могут фигурировать в качестве динамических переменных, $\ldots$ а
ВФ как носители информации не могут фигурировать в ее
аппарате$\gg$ [1, стр.~16]. Эти проблемы: отсутствие формальных
конструкций, описывающих распределения координат, и ограниченные
возможности наблюдения ввиду возможного появления пар и других
эффектов квантового поля (КП).

Первая из этих проблем~--- это математическая проблема
существования и единственности  плотности  вероятности координат
РКЧ,  которая обладала бы  известными формальными свойствами,
требуемыми квантовой механикой (КМ)  и специальной теорией
относительности (СТО). Ее хрестоматийное исследование приводит к
негативному ответу, [1,\linebreak стр.~16]: эти требования
несовместны. Здесь показывается, что проблема имеет исчерпывающее
положительное решение, если полный набор наблюдаемых координат
расширить с 3 пространственных до 4
прост\-ра\-нс\-т\-ве\-н\-но-временных, лишив время функции
динамического параметра распределения и формально уравняв его с
другими координатами. Такое представление плотности предъявляется,
и доказывается его единственность. Из последнего следует: если
физический постулат КМ о существовании вероятностного
распределения координат, обладающего плотностью, справедлив и для
РКЧ, то оно должно описываться найденной математической
конструкцией. Содержание такого распределения: вероятность событий
в пространстве-времени вместо позиций в пространстве. Изменение
смысла соответствует логике СТО и предсказывает новое свойство
РКЧ. Вместо стохастического танца, описываемого потоком
вероятности, имеем распределение случайных появлений частицы в
пространстве-времени при многократном воспроизведении системы.

Алгоритмы вторичного квантования позволяют это новое свойство РКЧ
доопределить на квантовое поле (КП) в виде характеристик
распределения частиц в пространстве-времени в дополнение к таковым
для импульсных состояний. Получены операторы этих распределений
для  свободных КП, а также~---  релятивистски инвариантное
описание КП в рамках предлагаемой модели. Лишение времени функции
динамического параметра и формальное уравнивание его с другими
координатами в статистической части описания РКЧ вносит и
некоторые новые элементы в понимание распределений энергии
-импульса. Исчерпывающий ответ на вопрос о соответствии данной
модели физической реальности могут дать только экспериментальные
исследования. Но некоторые ее теоретические свойства и следствия
позволяют надеяться на положительный ответ. Описание
статистической части РКЧ в терминах событий в пространстве-времени
соответствует описанию динамической части и логике СТО.  РКЧ и КП
естественным образом, без специальных исключений, вписываются в
систему общих статистичеких постулатов КМ. Подтверждаются для РКЧ
отношения неопределенностей Гейзенберга; при нынешнем состоянии
вопроса они не имеют  теоретического основания, поскольку
необходимый для их вывода закон распределения координат
отсутствует. Описание КП становится в некоторых деталях более
лаконичным и менее противоречивым: нет необходимости рассматривать
КП как смесь двух сортов частиц с присущими каждому сорту
операторами Фока, частицы и античастицы оказываются разными
состояниями одной частицы; энергия вакуумного состояния бозонного
КП равна нулю вместо парадоксальной бесконечности.

Способ решения проблемы  эффектов КП, предлагаемый  и
обосновываемый здесь: погружение наблюдения единичной РКЧ в
наблюдение КП. При этом сеансы с появлением большего числа частиц
и их отсутствием не должны учитываться. Теоретически этот прием
делает  возможной точную фиксацию координат, хотя и не для всех
типов частиц. В этом смысле и фотон имеет координаты. Для других
типов сохраняет силу нижний порог неопределенности координат
единичной РКЧ, обусловленный недопущением эффектов КП. Рассмотрены
теоретически возможные процедуры наблюдения. Но на пути к
эксперименту необходимы  дальнейшие исследования в русле теории
квантовых измерений.

\section*{2. Нерелятивистская частица}

\hspace{0.5cm} Поскольку статистическая часть достаточно
формализована только для нерелятивистской КМ, то начнем с фиксации
необходимых здесь ее деталей, ограничившись бесспиновой квантовой
частицей (КЧ). Пусть $\mathbf{x}=\{x^i\}=(x^1,x^2,x^3)$~--- вектор
пространственных координат частицы, элемент соответствующего
эвклидова пространства $\mathbf{E}$. Комплексная ВФ  $\psi
(\mathbf{x})$ определена на $\mathbf{E}$ и рассматривается как
элемент гильбертова пространства $\mathbf{H}$ с нормой
$L_2(\mathbf{E})$ и соответствующим произведением
$(\psi_1,\psi_2)=\int\limits_\mathbf{E}\psi_1(x)\psi^*_2(x)d\mathbf{E}$,
$d\mathbf{E}$~--- элементарный объем $\mathbf{E}$, $^*$~--- значок
комплексного сопряжения. Как правило, ВФ финитна в некотором кубе
$\mathbf{V}\subset \mathbf{E}$:  $\psi (\mathbf{x})=0$,
$\mathbf{x}\notin \mathbf{V}$, и этот несобственный интеграл
определен элементарно. Эволюция ВФ $\psi (t,\mathbf{x})$ на
интервале $(0,T)$ описывает движение КЧ. Она удовлетворяет
уравнению Шредингера, причем $\|\psi (t,\mathbf{x})\|$~--- его
динамический инвариант, нормированный условием $\|\psi
(t,\mathbf{x})\|=1$.

{\em К процедуре наблюдения}. Полная совокупность наблюдаемых
$y=\linebreak =\{y_i\}$~--- либо вектор  $\mathbf{x}$, либо
импульс. Пусть определены операции реализации движения КЧ в
течение времени $T$ в тождественных физических условиях,
отвечающих данной эволюции ВФ $\psi (t,\mathbf{x})$. В каждой
реализации в каждый момент времени $t$ наблюдается значение
$y(t)$. Траектории $y(t)$ в каждом сеансе определены, а при их
повторении~--- случайны. Предполагается, что для каждой функции
$\psi (t,\mathbf{x})$ и каждого $t$ определена мера вероятности:
$P(Q/t)=P[Q,\psi (t,\mathbf{x})]$ на подмножествах $Q$
пространства наблюдаемых, и соответственно, средние $\langle
y\rangle \left[\psi (t,\mathbf{x}) \right]=\linebreak =\{\langle
y_i\rangle (\psi )\}$, соответствующие бесконечному набору
замеров. Но это не вполне точно, поскольку такая процедура
обременена проблемой ВФ после измерения (коллапс ВФ). Поэтому
будем придерживаться более громоздкого, но и более корректного
понимания замера: для каждой реализации задается единственный
момент времени $t$ и  наблюдается значение $y$.

Система $\ll$квантовая частица$\gg$ интерпретируется  как
динамическая система с состоянием $\psi (\mathbf{x})$. Ее эволюция
сопровождается стохастическим процессом в пространстве
наблюдаемых, который в каждый момент времени $t$ описывается мерой
вероятности $P(Q/t)=P[Q,\psi (t,\mathbf{x})]$. Но в свете
сказанного справедлива и иная интерпретация. Именно. Реализации
системы в  физических условиях, отвечающих данной эволюции ВФ
$\psi (t,\mathbf{x})$, не тождественны.   Им  в описанных условиях
наблюдения соответствуют  разные (единственные) значения $y$. При
их многократном повторении значения $y$ случайны, и определено
вероятностное распределение $P(Q)$. Система $\ll$квантовая
частица$\gg$, таким образом, рассматривается как совокупность ее
реализаций. Иными словами, КЧ есть совокупность динамически
подобных, статистически связанных динамических систем. И динамика,
и статистика определены ВФ $\psi (t,\mathbf{x})$ так, как описано
выше.  Здесь эти модели тождественны, но вообще, как мы увидим,
последняя модель является более общей и применима к РКЧ.

{\em К формальному описанию}. Средние суть эрмитовы квадратичные
формы (ЭКФ) в  $\mathbf{H}$:  $\langle y_i\rangle (\psi )=(\psi ,
L_i\psi )$, $L_i$~--- соответствующий оператор (оператор $i$-й
наблюдаемой). Если $y=\mathbf{x}$, то постулируется наличие
плотности вероятности $f(t,\mathbf{x})$  и следующая ее связь с
ВФ: $f(t,\mathbf{x})=|\psi (t,\mathbf{x})|^2$, и
соответственно,~--- мера вероятности: $P(Q/t)=P\left[Q,\psi
(t,\mathbf{x})\right]=\linebreak ={\displaystyle\int} |\psi
(t,\mathbf{x})|^2d\mathbf{E}$, $Q\in \mathbf{E}$. Если $y$~---
импульс, то постулируется равенство $\langle y\rangle (\psi
)=J(\psi)$, где правая часть~--- соответствующий динамический
инвариант волнового уравнения (ВУ), ЭКФ. В этом случае мера
вероятности $P(Q,\psi)$ на подмножествах $Q$ пространства
наблюдаемых выражается известным образом через компоненты
спектральных функций операторов $L_i$.

Имеются исключения из правила $\psi \in L_2(\mathbf{E})$, когда
норма не определена и не удовлетворяется условие
$P(\mathbf{E},\psi)=1$. Но определены относительные вероятности
$P(Q_1,\psi)/P(Q_2,\psi)$. Отметим также, что  в любом
фиксированном кубе $\mathbf{V}\subset \mathbf{E}$ ВФ есть элемент
гильбертова пространства $\mathbf{H}(\mathbf{V})$ с нормой
$L_2(\mathbf{V})$. Фиксируя здесь $\|\psi\|=1$, можно сохранить за
$P(Q,\psi)$, $Q\subset \mathbf{V}$, смысл меры вероятности
наблюдения позиций $\mathbf{x}$, а за $f(t,\mathbf{x})$~--- ее
плотности, при следующей поправке к процедуре наблюдения:
учитываются только сеансы, показывающие $\mathbf{x}\in
\mathbf{V}$.

\section*{3. Релятивистская частица. Общее описание модели}

\hspace{0.5cm} Пусть $E$~--- действительное псевдоэвклидово
пространство с координатным вектором
$x=\{x^{\alpha}\}=(x^0,\mathbf{x})$, $\mathbf{x}\in \mathbf{E}$;
$x^0=ct$, $t$~--- время, $c$~--- скорость света; метрический
тензор $e=\{e_{\alpha \beta}\}$~--- диагональный: $e_{00}=-1$,
$e_{ii}=1$, $i>0$; $E^{*}$~--- его комплексный аналог $c$ с
элементами $n$ и произведением
$u_1u_2=u^{*}_{1}{\mathbf{u}}_{2}-u^{0*}_{1}u^{0}_{2}$. На
конечном брусе $V=(0,cT)\times \mathbf{V}\subset E$ определена ВФ
$\psi (x)$, отображающая $V$ в многообразие $U\subset E^{*}$,
$u^{2}\geq 0$, характерное для каждого рассматриваемого типа
частиц. Она удовлетворяет соответствующему ВУ и рассматривается
как элемент гильбертова пространства $H(V)$ с нормой
$\|\psi(x)\|^2={\displaystyle\int\limits_V}u^2dE$, $u=\psi(x)$,
$dE=dx^0d\mathbf{E}$ (норму пространства $\mathbf{H}$ обозначаем
$\|\psi(t,\mathbf{x})\|$) и соответствующим произведением
$(\psi_1,\psi_2)$. В отличие от нерелятивистской КМ каждая ВФ
описывается квантовый процесс в целом, а не его динамические
состояния в фиксированные моменты. {\em РКЧ определим как КС,
которая исчерпывающе описывается ВФ} $\psi (x)$. Для каждой ВФ
определены тождественные физические условия, в которых наблюдается
РКЧ.

Система $\ll$квантовая частица$\gg$ интерпретируется применительно
к РКЧ как совокупность реализаций (см.~конец п.~2) со следующими
отличиями. Сеанс наблюдения (замер) теперь~--- это операция
реализации этой системы и наблюдение за ней, фиксирующее значение
наблюдаемой $y$. Предполагается, что для каждой ВФ определена мера
вероятности $P(Q,\psi)$, соответствующая бесконечному набору
замеров, и соответственно, среднее $\langle y\rangle (\psi(x))$.
Время исключается из описания результатов наблюдения в качестве
динамического параметра, но присутствует в составе аргумента $x$
и, возможно, наблюдаемой $y$. Исключение особой роли времени из
статистической части РКЧ (для динамической она исключена)~---
необходимый элемент внедрения логики СТО. Остальные условия п.~2
остаются в силе. Главное из них: существование и единственность
значения $y$ при замере в новых условиях. Мы обсудим его ниже
применительно к конкректным наблюдаемым. Интерпретация системы
$\ll$квантовая частица$\gg$ как динамической системы с состоянием
$\psi (\mathbf{x})$, сопровождаемой стохастическим движением
частицы, оказывается неприменима к РКЧ. Отсутствие стохастического
процесса не так уж неожиданно: $\ll$Можно думать, что будущая
теория вообще откажется от рассмотрения временного хода
процессов$\ldots$ Описание процесса во времени окажется столь же
иллюзорным, какими оказались классические траектории в
нерелятивистской КМ$\gg$ [1], стр.~16.

Частица и античастица рассматриваются здесь как разные состояния
одной и той же частицы. Допускается и наличие отрицательных частот
в разложении ВФ и, соответственно,~--- наблюдение частицы
(единственной) с разными значениями заряда в разных сеансах. Это
не противоречит квантовому закону сохранения заряда, выполняемому
в среднем, но может не допускаться внешними для КМ законами,
такими, как закон сохранения электрического заряда.

Обратимся к проблеме  эффектов КП. По-видимому, единственный
способ наблюдения единичной РКЧ при наличии эффектов КП: погрузить
это наблюдение в наблюдение КП и  выделить из результатов
последнего характеристики единичной РКЧ.  Описание КП, обоснование
и детали такого погружения  см.~ниже, п.п.~6, 6.6, 7. Определению
РКЧ как КС, которая описывается ВФ  $\psi (x)$, отвечает
подсистема системы $\ll$квантовое поле$\gg$, определяемая
следующим образом: учитываются только замеры КП, показывающие
появление единственной частицы (п.~6.5). В целом, требования к
сеансу укладываются в формулу: учитываются только сеансы,
показывающие единственное значение наблюдаемой. Т.\,е. нет
необходимости в полной мере использовать процедуры наблюдения КП.
Достаточно замечать и отфильтровывать недопустимые этой формулой
замеры.

{\em Формальное описание} воспроизводит таковое для
нерелятивистской КЧ со следующими отличиями. ЭКФ $\langle
y_i\rangle (\psi)=(\psi,L_i\psi)$, и соответственно, операторы
наблюдаемых определены в  $H$, а не в  $\mathbf{H}$. Оператор
$L_{i}$ определяет набор возможных значений наблюдаемых и их
вероятности на таких процессах (собственных). Характеристики $y$,
и соответственно, их средние, обладают релятивистскими
трансформационными свойствами. $P(Q,\psi)$~--- релятивистский
инвариант. Последнее следует из равенства  $\langle y\rangle
(\psi)={\displaystyle\int} ydP(\psi)$, поскольку $\langle y\rangle
(\psi)$ и $y$ имеют одинаковую тензорную размерность. Размерность
плотности вероятности $g(y,\psi)$ определяется равенством
$dP=gdY=inv$, где $dY$~--- элементарный объем пространства
наблюдаемых. Далее следует детализация модели применительно к
конкретным наблюдаемым и типам частиц.

\section*{4. Распределение координат}

\hspace{0.5cm} Зафиксируем необходимые формальные свойства
плотности вероятности РКЧ. Как и все плотности средних величин в
КМ она выражается только непосредственно через значения ВФ и ее
1-х производных и квадратична. Кроме того, она определена на
полном наборе наблюдаемых, положительна при любых $\psi (x)$ и
имеет соответствующие трансформационные свойства. Подобные
математические конструкции назовем {\em допустимыми
представлениями плотности}. Поиск последних  в традиционном
направлении с пространственными координатами в качестве полного
набора, показал несовместность  трансформационного свойства
(временная компонента 4-вектора) с остальными,  и привел к выводу
об отсутствии соответствующего информативного свойства ВФ, [1].
Оказывается, допустимое представление плотности существует.
Проблема его существования (и единственности) имеет  положительное
решение, если традиционный полный набор наблюдаемых координат
расширить с 3 пространственных до 4 пространственно- временных,
лишив время функции динамического параметра распределения и
формально уравняв его с другими координатами. Такое уравнивание
соответствует логике СТО и свойствам ВУ. Полный набор наблюдаемых
расширяется до: $y=x$; результат каждого замера~--- фиксация
события: частица появилась в точке $x$ пространства-времени $E$.

{\bf{\em Теорема}}. Допустимое представление плотности существует
и единственно. Это:
\begin{equation}\label{1b1}
g(x)=\psi^2(x),\quad \|\psi
(x)\|^2=1.
\end{equation}
Действительно.  Элементарный объем пространства наблюдаемых
$dY=\linebreak =dE$ есть инвариант, и согласно  равенству
$dP=gdY=inv$, тензорная размерность плотности~--- скаляр.  Правая
часть (\ref{1b1}) также инвариантна и удовлетворяет остальным
требованиям допустимости. Это единственная такая  конструкция.
Имеется еще лишь один квадратичный инвариант, составленный из
компонент тензора $grad \psi (x)$. Но он не обладает свойством
положительности при любых $\psi (x)$. Это будет продемонстрировано
ниже (п.~4.1).

Зная  $g(x)$, можно найти плотность $g_1(\mathbf{x})$  вероятности
позиций $\mathbf{x}$,\linebreak $g_0(x^0)$~--- времени $x^0$,
пространственных координат в фиксированный момент как условную
$g_1(\mathbf{x}/x^0)$:
\begin{equation}\label{2b1}
\quad g_1(\mathbf{x})=\int\limits_{(0,\infty)}g(x)dx^0,\quad
g_0(x^0)=\int\limits_\mathbf{E} g(x)d\mathbf{E},\quad
g_1(\mathbf{x}/x^0)=g(x)/g_0(x^0).
\end{equation}

Доопределим на РКЧ постулат КМ о существовании  распределения,
обладающего плотностью. Тогда оно описывается (\ref{1b1}). Его
содержание: вероятность событий в пространстве-времени вместо
позиций в пространстве. Изменение смысла соответствует логике СТО
и предсказывает новое свойство КЧ. Как видим, из указанного общего
физического постулата следует конкретное математическое описание
(\ref{1b1}). А из последнего,~--- существенная деформация образа
КЧ при переходе к релятивистской теории. Вместо стохастического
танца частицы в пространстве, описываемого потоком вероятности,
имеем вероятностное распределение случайных появлений частицы в
пространстве-времени при многократном  воспроизведении  системы.
Конкретизируем (\ref{1b1}) применительно к некоторым частицам.

{\bf 4.1. Скалярный бозон}. Пространство $U$  одномерное,
действительное либо комплексное,  плотность:
$g(x,\psi)=|\psi(x)|^2$. Продолжим доказательство Теоремы. Имеем в
этом случае:
$$
grad \psi (x)=\left(\partial \psi (x)/\partial
{\mathbf{x}},\partial \psi (x)/\partial x^{0}\right)\in E;
$$
$$
\left(grad \psi (x)\right)^{2}=\left(\partial \psi (x)/\partial
{\mathbf{x}}\right)^{2}-\left(\partial \psi (x)/\partial
x^{0}\right)^{2}.
$$

Последняя конструкция есть упомянутый единственный, кроме
(\ref{1b1}), квадратичный инвариант. Существование ВФ, на которой
его значение отрицательно хотя бы при некоторых $x$ (и даже при
всех), очевидно. Это, например, $\psi (x)=\psi (x^{0})$. Для РКЧ,
обладающих дополнительными степенями свободы, существование такой
ВФ тем более очевидно. Теорема доказана.

{\bf 4.2. Векторный бозон}. ВФ есть вектор  $u(x)$, отображающий $E$
в  $E$, либо в его комплексный аналог   $E^*$. Имеем:
$u^2=\mathbf{u}^2-(u^0)^2$. Эвклидово пространство $U$
выделяется условием:  $u^2\geq 0$. Для последнего необходимо и
достаточно: $u^{\prime 0}=0$  в какой–либо фиксированной системе
отсчета  $x'$. Пространство $U$  тем самым оказывается определено
с точностью до преобразования  $u'\to u$, т.\, е.~--- до вектора
скорости  $\mathbf{v}$  системы  $x$ относительно  $x'$. Для
{\em массивного бозона} такая система $x'$   существует.
Это~--- его система покоя. Тем самым вектор  $\mathbf{v}$
фиксирован и пространство $U$  есть 3 мерное эвклидово сечение
псевдоэвклидового пространства  $E$. Плотность:  $g(x,\psi )=u^2(x)$.

{\em  Безмассовый бозон}, {\em фотон}, не имеет системы покоя, но
отсюда не следует отсутствие системы  $x'$,  $u^{\prime 0}=0$.
Более того, в отличие от массивного бозона, если такая система
$x'$  существует, то она не единственна. Действительно, выделим
случай, когда поле есть плоский волновой пакет с волновыми
векторами компонент, параллельными  вектору  $\mathbf{k}$. Включив
условие Лоренца в состав уравнений поля, и  предполагая, что
вектор $\mathbf{v}$ параллелен  $\mathbf{k}$, получим:  $u^{\prime
1}=0$. Легко убедиться, что  $u^0=u^1=u^{\prime 1}=u^{\prime
0}=0$, т.\,е. пространство $U$ есть поперечная плоскость с базисом
$u^2$, $u^3$  независимо от $|\mathbf{v}|$. То есть, калибровка
$u^0=u^1=0$ инвариантна на подгруппе группы Лоренца
$\mathbf{v}\uparrow \uparrow \mathbf{k}$. На основании сказанного
представляется естественным доопределить фотон постулатом:
существует система  $x'$,  $u^{\prime 0}=0$. Он исключает
неопределенность градиентно-неинвариантного выражения
$g(x)=u^2(x)$ и обеспечивает его положительность. Доопределяется
на фотон (\ref{1b1}) и непротиворечиво минимизируется различие
свойств бозонов: система покоя для фотона отсутствует, но ее
свойство $u^{0}=0$ сохраняет целое семейство систем. Но
достигается это ценой отказа от принципа градиентной
инвариантности электродинамики. Последний подтвержден ее опытом
(за единственным известным исключением). Но весь он связан со
значениями и распределениями напряженностей, энергии, импульсов,
моментов и не касается распределения координат фотонов. Только
эксперимент может определить альтернативный выбор: либо данный
принцип здесь неприменим, и справедливо $g(x)=u^2(x)$, либо
распределение координат фотона не определено.

{\bf 4.3. Электрон}. Пространство $U$ 4-мерное, комплексное с
элементами $u=\{u^{\alpha}\}$,  $\alpha =1,2,3,4$.
Трансформационные свойства элементов~--- спиноры, а $u^2$~---
временная компонента вектора. ВФ $u(x)=u(x^0,\mathbf{x})$
рассматривается обычно как траектория в гильбертовом пространстве
$\mathbf{H}(\mathbf{V})$. Она удовлетворяет ВУ Дирака, причем
$\|u(x^0,\mathbf{x})\|$~--- его динамический инвариант. Эти
свойства дают основания для равенства: $u^2(x)=\linebreak
=g(\mathbf{x}/x^0)$, $\|u(t,\mathbf{x})\|^2=1$, \cite{bib2}.
Введем новую ВФ $\psi(x)$, такую что $g(x)=\linebreak
=g_0(x^0)u^2(x)=\psi^2(x)$, $\|\psi(x)\|^2=1$. В силу ВУ она
совпадает с $u(x)$ c точностью до нормировки. Имеем:
$g_0(x^0)={\rm const}=1/cT$, $\psi=(cT)^{-1/2}u$,
$\|\psi(x)\|^2=1$. При $T\to \infty$, $V\to E$ предел
$\|\psi(x)\|^2$ здесь определен, если определен аналогичный
несобственный интеграл по  $\mathbf{E}$. Плотности $g(x)$,
$g(\mathbf{x}/x^0)$ совпадают с точностью до нормировки.

{\bf 4.4. К процедурам наблюдения}. Закон сохранения количества
частиц для РКЧ, вообще, отсутствует, и соответственно,~--- не
выполняются требования существования частицы (единственной) на
всем временном интервале и определенности траектории
${\mathbf{x}}(t)$ в каждой реализации. В каждый  момент времени
возможно    появление нескольких частиц (значений
${\mathbf{x}}(t)$), либо их отсутствие. Первое явление есть эффект
КП. Второе~--- отвечает определению РКЧ как КС, которая
описывается ВФ $\psi (x)$. В частности, допускается  как
существование частицы в течение всего времени $T$, так и ее
существование на пренебрежимо малом интервале (мгновенное
появление-исчезновение).

Рассмотрим сначала  последний случай (время жизни частицы
пренебрежимо мало). В каждой реализации наблюдается единственное
событие $y=x$. Содержание замера:  создание физических условий,
таких, что наблюдаемое событие будет зафиксировано. Фиксация
значения $y=x$ производится мгновенно, но момент времени $t$ не
задан, а определяется вместе с другими координатами. Пусть
прибор~--- электронный микроскоп; пучок электронов
$\ll$прощупывает$\gg$ пространство, и в некоторый момент
происходит столкновение с частицей-объектом; позиция частицы
фиксируется точкой на экране,~--- следом единственного рассеянного
электрона. Для полной фиксации события необходима также фиксация
момента столкновения. (\ref{1b1}) связывает  плотность вероятности
этих событий с ВФ. Здесь не только в результатах, но и в замерах,
время формально уравнено с другими координатами. Если фиксация
времени $t$ в замерах затруднительна, то можно ограничиться
наблюдением плотности пространственных координат
$g_{1}({\mathbf{x}})$ и проверять (\ref{1b1}), используя
(\ref{2b1}). Здесь деформация образа КЧ при переходе к
релятивистской теории особенно наглядна. В нерелятивистской КМ
эволюция ВФ сопровождается стохастическим процессом, описываемым
потоком вероятности. Здесь же мы имеем вероятностное распределение
единичных  появлений частицы в пространстве-времени при
многократном  воспроизведении системы в целом, которое не сводится
к стохастическому движению частицы в пространстве.

В общем случае для того, чтобы обеспечить единственность событий в
реализациях, требуются дополнительные условия даже при отсутствии
эффектов КП. Поэтому в каждом  сеансе наблюдения будем задавать
момент времени $t$ подобно п.~2, и  наблюдать значение
${\mathbf{x}}(t)$. Если эффекты КП отсутствуют, то это значение (и
событие $x=\left(ct,{\mathbf{x}}(t)\right)$ либо существует и
единственно, либо отсутствует. Результаты замеров упорядочиваются
следующим образом. Разобьем брус $V\subset E$ на брусы $v(\xi )$,
каждый помечен значением $x=\xi =(\xi^{0},\xi )$, принадлежащим
ему. Сеансы проводятся сериями, в которых $t$ пробегает все
значения $\xi^{0}$. Подсчитывается распределение количеств $n(\xi
)$ показаний $x\in v(\xi )$. Пусть число $N=\Sigma n(\xi )$
положительно при достаточно большом количестве серий, начиная с
определенного этапа предельного перехода $diam\,v(\xi )\to 0$,
$\forall \xi$. При бесконечном повторении серий и соответствующей
нормировке получаем  плотность вероятности $g(x)$. (\ref{1b1})
связывает эту плотность с ВФ. Как видим, вообще, особая роль
времени неустранима из описания процедуры наблюдения, но
устраняется из описания его результатов, $g(x)$.

\section*{5. Распределение энергии-импульса}

\hspace{0.5cm} Пусть РКЧ~---  скалярный бозон.  Наблюдаемая~---
вектор 4-импульса   $p=(p^0,\mathbf{p})\in E$. Его среднее
традиционно записывается как динамический инвариант
$J=\{J^{\alpha}\}$, ЭКФ в $H$:
$$
J^{\alpha}(\psi )={\displaystyle\int}
\Pi^{0\alpha}(x^{0},{\mathbf{x}})\,d{\mathbf{E}}=J^{\alpha}(x^{0})={\rm
const},
$$
где $\Pi^{0\alpha}$~--- соответствующие компоненты псевдотензора
энергии-импуль\-са. В релятивистской системе единиц $\hbar =c=1$:
$$
\Pi^{00}=\partial_{0}\psi^{*}\partial_{0}\psi
+\partial_{i}\psi^{*}\partial_{i}\psi +m^{2}\psi^{*}\psi ;\quad
{\bf{\Pi}}^{0i}=\partial_{0}\psi^{*}\partial_{i}\psi
+\partial_{0}\psi \partial_{i}\psi^{*}.
$$

Его собственные ВФ образуют семейство: $\psi_{k}=r_{k}\exp
(ip_{k}x)$, $k$ пробегает значения $\ldots ,-2,-1,1,2,\ldots$,
${\bf p}_{k}$ пробегает известную 3-мерную решетку, $p^{0}_{\bf
k}=\varepsilon_{k}sign\,(k)$, $\varepsilon_{k}=({\bf
p}^{2}_{k}+m^{2})^{1/2}$,
$r_{k}=\left(2|V|\varepsilon_{k}\right)^{-1/2}$,
$|{\mathbf{V}}|$~--- объем куба ${\bf V}$, (нормировка
$\ll$частица в единичном объеме$\gg$). Распределение описывается в
терминах средних количеств частиц $n_{k}$ с данным 4-импульсом
$p_{k}$ в качестве наглядного полуфабриката квантового поля. А
точнее, $n_{k}$~--- среднее количество замеров с результатом
$p_{k}$. Заменяя нормировку на $r_{k}=\left(2|V|\varepsilon_{\bf
k}\right)^{-1/2}$ (частица в единичном 4-объеме), получаем более
симметричное описание:
$$
J^{\alpha}(\psi )={\displaystyle\int} \Pi^{0\alpha}(x)\,dE.
$$
Здесь $\{\Pi^{0\alpha}\}$~--- вектор, а не псевдовектор. В
пространстве $l_{2}$ коэффициентов $C=\{C_{k}\}$ разложения $\psi
=\Sigma_{k}C_{k}\psi_{k}$: $n_{k}=|C_{k}|^{2}$. А при
дополнительном условии $\|C\|=1$  это~--- безусловное
распределение вероятности появлений единичной частицы в
пространстве 4-импульсов. Каждый замер должен фиксировать
единственную точку в этом пространстве. Как правило, это выполнимо
в силу относительного постоянства импульса. При этом, координаты,
как и продолжительность  фиксации, не присутствуют явно, а
ограничены только условием  $x\in V$.  Определен предел  вектора
$J$ при $V\to E$ (соответствующий несобственный интеграл). При
таком предельном переходе условие $x\in V$ также исключается из
процедуры наблюдения. В каждом замере, таким образом, частица
имеет фиксированный 4-импульс и неопределенное положение в
пространстве-времени.   При многократном воспроизведении системы
значение 4-импульса случайно, и определено его среднее значение
$J=\{J^{\alpha}\}$ и вероятностное распределение  $P(Q)$. Система
$\ll$релятивистская квантовая частица$\gg$, таким образом,
рассматривается как совокупность ее реализаций. Она, вообще,
несводима к модели стохастического процесса, поскольку время
появления частицы в каждой реализации не определено. Наглядный
пример: короткоживущая частица (п.~4), рассматриваемая на этот раз
в пространстве 4-импульсов.

Среднее значение 4-импульса $J=\{J^{\alpha}\}$  не зависит от $T$.
Это значит, в частности, что с точностью до нормировки,
безусловное распределение совпадает с условным  в любой момент
времени  $t$. Последнее, соответственно, не зависит от    $t$. В
этом статистический смысл законов сохранения. Хрестоматийным
распределениям приписывается именно смысл условных. Но этот смысл
порождает известное противоречие: он требует мгновенной фиксации
импульса при замерах, тогда как ограничения  точности наблюдения
РКЧ требуют, вообще, продолжительной фиксации. В случае
комплексной ВФ к энергии и импульсу добавляется заряд, а
многокомпонентной~--- спин.

\section*{6. Квантовое поле. Заполнение пространства-вре\-ме\-ни}

\hspace{0.5cm} {\bf 6.1.Общее описание}. Новые свойства РКЧ
доопределяются на КП. Адекватная база для этого: концепция КП как
системы тождественных частиц, \cite{bib2}.  Рассмотрим систему
тождественных невзаимодействующих частиц. Пусть $y=\{y_k\}$~---
наблюдаемая совокупность характеристик одной КЧ и $Y=\{Y_k\}$~---
системы. Пусть $\{\psi_i\}$~--- собственный базис некоторой
физической величины в $H$; $n=\{n_i\}$~--- набор его чисел
заполнения, $n_i=0,1,2,\ldots ,N$; $\Phi(n)$~--- ВФ системы,
выраженная через $n$, симметризованная. Ограничимся рассмотрением
только бозонных полей.

ВФ $\Phi(n)$~--- элемент гильбертова пространства $l_{2}$ с
произведением  $(\Phi_1,\Phi_2)$, нормированный:
$\|\Phi(n)\|^2=1$. Конкретизация пространства хрестоматийна (см.,
например, \cite{bib2}, стр.~311~--~314),  но здесь она не
понадобится. Определены операции реализации системы в
тождественных физических условиях, отвечающих данной  ВФ. Над
каждой реализацией проводится единственный сеанс наблюдения
(замер), фиксирующий значение $n$. Как и единичная РКЧ, система
рассматривается как совокупность таких динамически подобных,
статистически связанных реализаций. Время исключается из описания
результатов наблюдения в качестве динамического параметра (но
вообще,~--- не из процесса наблюдения). При повторении замеров
количество частиц $N=\Sigma_{i}n_{i}$ меняется, принимая значения
$0,1,2,\ldots$ Т.\,е. имеет место исчезновение и появление частиц
от замера к замеру. Это соответствует модели КП в рамках
корпускулярной концепции. Для каждой ВФ   определены их средние
$\langle Y\rangle$, соответствующие бесконечному набору замеров.
Не вдаваясь в детали процесса измерения, сосредоточимся на
формальном описании так определенного КП. Отметим лишь, что и
здесь остается актуальной проблема коллапса ВФ, и повидимому,
применимы детализации замеров п.~4.4 с соответствущим обобщением.
Средние суть ЭКФ: $\langle Y\rangle (\Phi)=(\Phi,\Lambda\Phi)$,
$\Lambda$~--- соответствующие операторы. Техника вторичного
квантования полностью воспроизводит применительно к РКЧ
нерелятивистский аналог \cite{bib2}, \cite{bib3} со сделанной выше
оговоркой о роли времени. Определяются характеристики безусловных
распределений $n$, а не для фиксированных значений $t$. Введем
операторы уничтожения и рождения частицы $a_{i}$, $a^{+}_{i}$ и
таковые, отнесенные к точке $x$~--- волновые операторы (ВО):
\begin{equation}\label{3b1}
\Psi(x)=\Sigma_i\psi_i(x)a_i,\quad
\Psi^+(x)=\Sigma_i\psi^*_i(x)a^+_i.
\end{equation}

Для синтеза операторов $\Lambda =\{\Lambda_{k}\}$ характеристик
системы $Y=\{Y_k\}$ записываем среднее для единичной частицы в
пространстве $l_{2}$ коэффициентов $C=\{C_{k}\}$ разложения $\psi
=\Sigma_{k}C_{k}\psi_{k}$, и производим замену:
\begin{equation}\label{4b1}
\begin{array}{c}
\langle y\rangle (\psi^{*},\psi
)=\Sigma_{ij}l_{ij}C^{*}_{i}C_{j},\quad
L_{ij}=(\psi_{i},L\psi_{j});\\[2mm]
C_{j}\to a_{j},\quad C^{*}_{i}\to a^{+}_{i},\quad \Lambda
=\Sigma_{ij}l_{ij}a^{+}_{i}a_{j}.
\end{array}
\end{equation}
Либо, используя ВО:
\begin{equation}\label{5b1}
\begin{array}{c}
\langle y\rangle (\psi^*,\psi)=(\psi ,L\psi)={\displaystyle\int}
\psi^* (x)L\psi
(x)dE;\\[2mm]
\psi \to \Psi(x),\quad \psi^*\to \Psi^+(x),\quad \Lambda=\langle
y\rangle \left(\Psi^+(x),\Psi(x)\right) .
\end{array}
\end{equation}
Здесь в произведении ВО рассматриваются как элементы $H$.
Пусть\linebreak $\{\psi_i\}$~--- собственный базис оператора $L$,
а $P_i$~--- вероятности соответствующих значений  $y_i$ при
наблюдении единичной РКЧ  (или соответствующие средние количества
замеров). Тогда, не прибегая к ВО и к (\ref{4b1}), и производя
замену: $P_i\to n_i$, можно записать
\begin{equation}\label{6b1}
\Lambda =\Sigma_{i}n_{i}y_{i}.
\end{equation}
Операторы, характеризующие распределения координат частиц,
отсутствуют в РКТ, как и для единичной РКЧ, и (\ref{1b1}),
(\ref{4b1}) восполняют этот пробел.  Оператор $\Lambda (Q)$
количества частиц в области $Q\subset E$:
$$
\begin{array}{c}
\Lambda (Q)=\left(\Psi^{+}(x),f(x)\Psi
(x)\right)={\displaystyle\int\limits_Q}\Psi^{+}(x)\Psi
(x)\,dE=\Sigma_{ij}l_{ij}(Q)a^{+}_{i}a_{j},\\[2mm]
l_{ij}(Q)={\displaystyle\int\limits_Q}\psi^{*}_{i}(x)\psi_{j}(x)\,dE,
\end{array}
$$
где $f(x)=1$, $x\in Q$, $f(x)=0$, $x\notin Q$,
$\|\psi_{i}(x)\|^{2}=1$. Справедливость (\ref{6b1}), однако,
ограничена случаем, когда каждому наблюдаемому значению $y$
отвечает единственное событие $x$. Оператор $\Lambda
({\mathbf{Q}})$ количества частиц в области ${\mathbf{Q}}\subset
{\mathbf{E}}$ совпадает с оператором $\Lambda (Q)$,
$Q={\mathbf{Q}}\times (0,T)\subset E$.

Количество частиц в данном состоянии есть результат наблюдения, не
зависящий от выбора системы координат. Соответственно, набор чисел
заполнения релятивистски инвариантен, так же как и операции над
ним $a_{i}$, $a^{+}_{i}$. Поэтому ВО обладают релятивистскими
трансформационными свойствами ВФ частицы $\psi$, а операторы
$\Lambda$~--- свойствами их аналогов $L$.

{\bf 6.2. Представление квантового поля в числах заполнения
импульсных состояний}. Пусть  $\{\psi_i(x)\}$~--- собственный
базис энергии, импульса, спина и заряда. Применительно к этим
наблюдаемым (\ref{4b1}), (\ref{5b1}), (\ref{6b1}) дает
хрестоматийные операторы  со следующими поправками: частицы и
античастицы рассматриваются как разные состояния одной частицы, а
не как два сорта частиц с разными операторами уничтожения и
рождения (генезис и основополагающие работы такого хрестоматийного
понимания  см. \cite{bib4}, п.~1.2); отсутствует расходящееся
слагаемое в операторе энергии. Так, для энергии заряженного бозона
имеем, учитывая (\ref{4b1}):
\begin{equation}\label{7b1}
\begin{array}{c}
\langle y\rangle
(\psi^*,\psi)={\displaystyle\sum\limits_{k>0}}\varepsilon_{k}
(C^{*}_{k}C_{k}+C^{*}_{-k}C_{-k});\\[2mm]
\Lambda =
{\displaystyle\sum\limits_{k>0}}\varepsilon_{k}(a^{+}_{k}
a_{k}+a^{+}_{-k}a_{-k})={\displaystyle\sum\limits_{k>0}}
\varepsilon_{k}(n_{k}+n_{-k}),
\end{array}
\end{equation}
в отличие от хрестоматийного
\begin{equation}\label{8b1}
\Lambda ={\displaystyle\sum\limits_{k>0}}
\varepsilon_{k}(n_{k}+n_{-k}+1).
\end{equation}
Эти различия обусловлены разными моделями КП. Имеются две
хрестоматийные  модели КП:  1) нерелятивистская КС с координатным
вектором  $n$; 2) классическое поле, квантованное по правилам
нерелятивистской КМ. Добавим описанную здесь модель: 3) система
динамически подобных статистически связанных реализаций КП.
Согласно 1) КП есть стохастический процесс, описываемый ВУ
Шредингера $\partial \Phi (t,n)/\partial t=\linebreak ={\widehat
H}\Phi (t,n)$, а ВО (\ref{3b1})~---  оператор поля в
гейзенберговом представлении. Известное свойство таких операторов
КМ: знак $+$ или $-$  при $\varepsilon_{k}t$ означает,
соответственно, добавление или вычитание кванта энергии в
состоянии $\psi_{k}$ в процессе эволюции системы. В согласии с
этим в каждой из сумм (\ref{3b1}) должны появиться в
соответствующих позициях операторы как уничтожения, так и рождения
частицы, что обусловливает необходимость рассматривать систему как
смесь двух сортов частиц. Упорядочение последовательности этих
операторов в представлении оператора энергии с учетом их
коммутационных свойств и приводит к (\ref{8b1}). В модели 2)
классическое поле представляется как процесс колебаний системы
осцилляторов и квантуется по правилам нерелятивистской КМ, что
тоже приводит к (\ref{8b1}). Модель 3) характерна отсутствием
стохастического процесса, ВФ не зависит от времени, отмеченная
связь операторов Фока со знаком при $\varepsilon_{k}t$
отсутствует, и отпадает необходимость рассматривать систему как
смесь двух сортов частиц. Последовательность операторов Фока в
представлении оператора энергии не требует упорядочения, и
справедливо (\ref{7b1}).

Для энергии истинно нейтрального бозона имеем, учитывая
(\ref{4b1}):
\begin{equation}\label{9b1}
\begin{array}{c}
\langle y\rangle
(\psi^*,\psi)=(1/2){\displaystyle\sum\limits_{i>0}}\varepsilon_{i}
(C^{*}_{i}C_{i}+C_{i}C^{*}_{i})={\displaystyle\sum\limits_{i>0}}
\varepsilon_{i}C^{*}_{i}C_{i};\\[2mm]
\Lambda = {\displaystyle\sum\limits_{i>0}}\varepsilon_{i}a^{+}_{i}
a_{i}={\displaystyle\sum\limits_{i>0}} \varepsilon_{i}n_{i}.
\end{array}
\end{equation}
В отличие от моделей 1), 2) здесь нет нужды трансформировать
равенство $C^{*}_{i}=C_{-i}$  в  $a^{+}_{i}=a_{-i}$.   Правило
(\ref{4b1}) применяется непосредственно к последнему значению
$\langle y\rangle (\psi^{*},\psi )$. Его генезис, отраженный в
предшествующем значении, несущественен для (\ref{4b1}).

Оператор $\Lambda (Q)$ количества частиц в области $Q\subset E$
задан (\ref{4b1}), с отмеченной выше оговоркой: каждому
наблюдаемому импульсу отвечает единственное событие  $x$. Базис
$\{\psi_{i}(x)\}$ должен быть перенормирован:
$\|\psi_{i}(x)\|^{2}=1$, $r_{i}=|V|^{-1/2}$, $\psi_{i}=r_{i}\exp
(ip_{i}x)$, вместо $\ll$частицы в единичном объеме$\gg$. Имеем:
\begin{equation}\label{10b1}
\begin{array}{c}
\Lambda (Q)=\Sigma_{ij}l_{ij}(Q)a^{+}_{i}a_{j},\\[2mm]
l_{ij}(Q)= {\displaystyle\int\limits_{Q}}\exp
\left(i(p_{j}-p_{i})x\right) \,dE/|V|,\\[2mm]
l_{jj}(Q)=|Q|/|V|;
\end{array}
\end{equation}
$$
\Lambda
(V)=\Sigma_{ij}l_{ij}(V)a^{+}_{i}a_{j}=\Sigma_{i}n_{i}=N;\quad
l_{ij}(V)=0,\quad i\ne j;
$$
$$
l_{ij}(V)=1,\quad i=j.
$$

{\bf 6.3. От квантовой частицы к квантовому полю. Квантование или
клонирование?} Вернемся к анализу моделей КП.   С точностью до
исходных положений модель 3) непротиворечива. Модели 1) и 2)
приводят к противоречию наблюдаемого распределения энергии с
предсказаниями теории. В них $\ll$мы сталкиваемся с одной из
расходимостей, к которым приводит отсутствие полной логической
замкнутости существующей теории$\gg$, \cite{bib1} стр.~24.
Противоречие не сводится только к расходимости энергии. Если за
классический прототип принять конечную частичную сумму
соответствующего Фурье-разложения ВФ, то собственные значения
определены и должны наблюдаться в замерах согласно принципу
суперпозиции. Но наблюдается только энергия частиц. Допущение о
том, что энергия вакуума скрыта от наблюдения, противоречит этому
принципу, поскольку последний безоговорочно определяет
распределение именно наблюдаемых значений. Не спасает положения и
тот факт, что энергия определена с точностью до постоянного
слагаемого. Почему последнее различно для механического
осциллятора и осциллятора~--- компонента поля? Эти  трудности
преодолеваются волевым исключением энергии вакуума из собственных
значений энергии поля, но строгое обоснование этого приема
отсутствует. Непротиворечивость модели 3) делает ее, повидимому,
более адекватной физической реальности. Кроме того, этот факт
служит дополнительным теоретическим аргументом в пользу
представленного здесь понимания РКЧ и ее законов распределения.

Модели КП опираются, помимо общих постулатов КМ, на специальные
допущения. В 3) система РКЧ дополняется феноменом несохранения
количества частиц в замерах. В 2) предполагается наличие феномена
квантования не только в динамике материальных точек (более
обще,~--- классических механических систем), но и  полей тоже, и
вообще,~--- $\ll$квантуется все$\gg$. Эти допущения предполагают
разные алгоритмы и физическое содержание генезиса КП из КЧ:
повторное квантование строго по тем же законам, что и первое, либо
переход к системе тождественных частиц~--- $\ll$клонирование
частицы$\gg$. Вопреки хрестоматийному мнению, \cite{bib2}
стр.~316, результаты этих алгоритмов не тождественны, хотя и легко
уравниваются волевыми поправками. Если противоречивость модели 1)
можно объяснить ее $\ll$нерелятивистскостью$\gg$, то модели 2), не
исключено,~--- только отказом от предположения о наличии феномена
квантования  в динамике  полей.

{\bf 6.4. Представление характеристик квантового поля в числах
заполнения ячеек пространства-времени}. Введем в рассмотрение
собственный базис координат, для унификации с дискретным базисом
импульса~--- на допредельном уровне римановых интегральных сумм.
Разобъем брус $V\subset E$ на брусы $v(\xi)$ с объемом $w(\xi)$,
каждый помечен значением $x=\xi$, принадлежащим ему. Определим
семейство функций  $\psi(x,\xi)$  с параметром $\xi$:
$\psi^2(x,\xi)=w^{-2}(\xi)$, $x\in v(\xi)$; $\psi (x,\xi)=0$,
$x\notin v(\xi)$.

Рассмотрим единичную частицу с ВФ $\psi (x)$. Аппроксимируем ВФ
ступенчатой функцией $\psi'(x)=\psi (\xi)$, $x\in v(\xi)$. Функции
$\psi (x,\xi)$, $\psi'(x)$ суть элементы подпространства
$H'\subset H(V)$ конечной размерности с ортогональным базисом
$\psi (x,\xi)$ и, с точностью до аппроксимации:
\begin{equation}\label{11b1}
\begin{array}{c}
\left(\psi'_1(x),\psi'_2(x)\right)=\Sigma_{\xi} \psi^{\prime
*}_1(\xi)\psi'_2(\xi)w(\xi);\quad \|\psi'(x)\|^2=1,\\[2mm]
\|\psi (x,\xi)\|^2=w(\xi)^{-1};\\[2mm]
\psi'(x)=\Sigma_{\xi}\psi (\xi)\psi (x,\xi)w(\xi)\to
\psi(x)={\displaystyle\int\limits} \psi (\xi)\delta(x-\xi)d\xi ,\\
[2mm]
w(\xi)\to 0.\\[2mm]
P(Q,\psi'(x))={\displaystyle\int\limits_{Q'}}\psi^{\prime 2}(x)dE=
\Sigma_{\xi}P(\xi),\ P(\xi)=\psi^2(\xi)w(\xi).
\end{array}
\end{equation}
Здесь ${\bf Q}’$~--- минимальный набор брусов  $v(\xi)$,
покрывающий ${\bf Q}$.

Доопределим аппарат вторичного квантования на собственный базис
 $\psi (x,\xi)$: $n=\{n(\xi)\}$~--- множество чисел
заполнения брусов $v(\xi)$, $n(\xi)=0,1,2,\ldots ,N$. Производя
замену: $P(\xi)\to n(\xi)$, получим оператор количества частиц в
области $Q'\subset E$:
$$
\Lambda (Q')=\Sigma_\xi n(\xi),\ \xi : v(\xi)\in Q';
$$
$n(\xi)$ выполняют здесь роль собственных чисел этого оператора.

{\bf 6.5. Уравнения поля}. Оператор правой части ВУ бозона:
$B=\linebreak =\partial_{\alpha}\partial_{\alpha}-m^{2}$. Его
квантово-полевой аналог и уравнение КП:
\begin{equation}\label{12b1}
\quad B=\left(\Psi^{+}(x),B\Psi
(x)\right)=\Sigma_{ij}b_{ij}a^{+}_{i}a_{j},\quad
b_{ij}=(\psi_{i},B\psi_{j});\quad B\Phi (n)=0.
\end{equation}

Пусть $\{\psi_{i}(x)\}$~--- собственный базис энергии, импульса,
спина и заряда. Поскольку он удовлетворяет ВУ частицы, то
$B\psi_{i}(x)=0$; $b_{ij}=0$, $\forall i$, $j$. (12) выполняется
тривиально, не ограничивая выбор $\Phi (n)$. Если КП представлено
в числах заполнения ячеек пространства-времени, то под
$\partial_{\alpha}\partial_{\alpha}$ понимается конечно-
разностная аппроксимация 2-й производной ступенчатой функции
$\psi'(x)$ (п.~6.4).  ВУ (\ref{12b1}) есть нетривиальная
алгебраическая система линейных однородных уравнений (сравнить с
хрестоматийным $\partial \Phi (t,n)/\partial t={\widehat H}\Phi
(t,n)$, \cite{bib2} стр.~) с комплексной матрицей $B$.

{\bf 6.6. Единичная частица как подсистема квантового поля}.
Рассмотрим единичную частицу с ВФ $\psi(x)$  в терминах вторичного
квантования как подсистему  $N=1$ системы $\ll$КП$\gg$,
определяемую следующим образом: при наблюдении учитываются только
замеры КП, показывающие $N=1$. Имеем: $n = (0,0,\ldots
0,1,0,\ldots )$, $n_i = 0,1$. Термин $\ll$тождественные
частицы$\gg$ теряет смысл, и ВФ $\Phi(n)$  не содержит оператора
перестановки. Найдем среднее количество частиц в области $Q\subset
\linebreak \subset E:\langle N\rangle
(Q,\Phi)=\left(\Phi,\Lambda'(Q)\Phi\right)$, пользуясь
конечномерной аппроксимацией, п.~6.4. $\{\psi(x,\xi)\}$~---
собственный базис наблюдаемой. Каждому $n$ отвечает событие
$x(n)\in E$. Отождествим точку $x(n)$ с пометкой $\xi$ включающего
ее бруса  $v(\xi)$. Имеем: $\Phi(n)=\psi(x(n))$. В согласии с
определением ВФ $\psi (x)$ (п.~3) следует положить
$(\Phi_1,\Phi_2)=\Sigma_n\Phi_1(n)\Phi_2(n)w(\xi=\linebreak
=x(n))$. Имеем: $\|\Phi(n)\|^2=1$; $\langle N\rangle
(Q,\Phi)=\left(\Phi,\Lambda'(Q)\Phi\right)=\Sigma_n
\Phi^2(n)=\linebreak =\Sigma_{\xi} \psi^2(\xi)w(\xi)=P(Q',\psi)\to
{\displaystyle\int\limits_Q} \psi^2(\xi)\,d\xi$, $w(\xi)\to 0$;
$n:x(n)\in Q$, $\xi:v(\xi)\in Q'$.

Таким образом, волновые функции $\Phi(n)$ и $\psi (x)$ суть
тождественное описание данной подсистемы КП, а ее наблюдение дает
тот  же результат, что и описанное выше непосредственное
наблюдение единичной частицы.

\section*{7. Об отношениях неопределенностей и оценках точности
наблюдения}

\hspace{0.5cm} В нерелятивистской КМ отношения неопределенностей
Гейзенберга суть следствие статистических постулатов, приведенных
выше. В рамках традиционной модели релятивистской КЧ они уже не
имеют этого теоретического основания, поскольку необходимый для
этого закон распределения координат отсутствует. Тем не менее, они
используются в том же виде, строго говоря, уже в качестве
самостоятельного постулата. Здесь эти отношения получают
обоснование. Их легко получить, воспроизводя хрестоматийный вывод,
например, \cite{bib3} стр.~ 66~--~68, применительно к $\psi (x)\in
H$. При этом, если в нерелятивистской КМ отношения
координаты-импульс и время-энергия выводятся разными способами и
имеют разный смысл \cite{bib3}, стр.~ 185~--~188, то здесь они
обладают полной формальной и смысловой симметрией. Проблема
эффектов КП при наблюдении единственной РКЧ решается здесь
погружением его в наблюдение КП. При этом сеансы с появлением пары
и большего числа частиц не учитываются. Обсудим в этой связи
теоретический нижний порог неопределенности координат РКЧ:
$\triangle x>\triangle x_{\min}=\hbar c/\varepsilon$
($\varepsilon$~--- энергия), обусловленный этими отношениями и
недопущением эффектов КП. Он теряет  силу применительно к истинно
нейтральным частицам, и к тем заряженным частицам, для которых
закон сохранения заряда имеет квантовый характер, будучи выполняем
только в среднем. Тогда возможно  наблюдение частицы
(единственной) с разными значениями заряда в фиксированной паре
сеансов. Порог заменяется минимальным  практически приемлемым
значением вероятности фиксации единственной частицы при наблюдении
КП. Теоретически возможна точная фиксация координат в замере. В
этом смысле и фотон имеет координаты. Нарушение порога может не
допускаться внешними для КМ законами, такими, как всеобщий закон
сохранения электрического заряда. В этом случае появление в
замерах единичной РКЧ противоположного заряда невозможно, как и
экспериментальное воспроизведение ВФ с наличием отрицательных
частот в разложении. Таков электрон.

Требует специального осмысления отношение неопределенностей
время~---~скорость~---~импульс $\mathbf{v}\triangle t\triangle
p>\hbar$ и вытекающая из него связь точности наблюдения импульса с
продолжительностью сеанса наблюдения   $\triangle p>\hbar
/c\triangle t$. Прежде всего, имеет ли смысл понятие $\ll$скорость
КЧ$\gg$ в данной модели, и если да, то какой?

\section*{8. Выводы}

\hspace{0.5cm} Показано, что при достаточно полном включении
статистической части РКЧ в логику СТО информативные свойства ее ВФ
обретают объем, сравнимый с нерелятивистской КМ. Появляется,
вопреки хрестоматийному мнению, теоретическая возможность
наблюдения координат РКЧ, и определено представление их плотности
вероятности $g(x)=\psi^{2}(x)$, отвечающее всем формальным
требованиям КМ и СТО. Здесь ВФ $\psi (x)$ отображает
пространство-время $E$ в эвклидово подпространство $U$,
характерное для каждого из рассмотренных типов частиц: бозона,
действительного и комплексного, скалярного и векторного, включая
фотон, электрона. Плотность $g(x)$ в простейших случаях формально
подобна плотности вероятности координат нерелятивистской частицы,
но имеет иное содержание: иные условия наблюдения, иные свойства,
иной смысл. Сеансы наблюдения не обусловлены моментом времени $t$,
т. е. время исключается из описания распределений в качестве
динамического параметра, но присутствует в составе аргумента ВФ и
наблюдаемой $y=x$; соответственно, плотность теперь~---
релятивистский инвариант; вместо стохастического танца частицы,
описываемого потоком вероятности с пространственной плотностью
$f(t,{\mathbf{x}})=|\psi (t,{\mathbf{x}})|^{2}$, имеем
вероятностное распределение появлений частицы в
пространстве-времени при многократном воспроизведении  системы.
Применительно к фотону эта формула обусловлена дополнительным
постулатом, альтернативным принципу градиентной инвариантности
электродинамики. В рамках подобной модели уточняется  также смысл
распределений энергии-импульса РКЧ. Отношения неопределенностей
Гейзенберга суть следствие постулатов нерелятивистской КМ. В
рамках традиционной модели РКЧ они уже не имеют этого
теоретического основания, поскольку необходимый для этого закон
распределения координат отсутствует. В рассматриваемой модели эти
отношения получают обоснование. Предложена и теоретически
обоснована процедура наблюдения частицы, фильтрующая эффекты
квантового поля КП. Соответственно, снимаются ограничения точности
наблюдения РКЧ основанные на недопущении этих эффектов. Новые
свойства РКЧ доопределяются на КП в виде характеристик
распределения частиц в пространстве-времени в дополнение к таковым
для импульсных состояний. Получены операторы этих распределений
для свободных КП. В рамках этой модели обретают новые свойства и
распределения частиц на импульсных состояниях. Отпадает
необходимость рассматривать КП как смесь двух сортов частиц с
присущими каждому сорту операторами Фока. Частицы и античастицы
оказываются разными состояниями одной частицы. Оператор энергии
бозонного КП не содержит расходящегося слагаемого, и энергия
вакуумного состояния равна нулю вместо парадоксальной
бесконечности. В арсенале РКТ имеется два алгоритма генезиса КП из
КЧ с разным, вообще, физическим содержанием: повторное квантование
строго по тем же формальным законам, что и первое, либо переход к
системе тождественных частиц ($\ll$клонирование частицы$\gg$).
Согласно представленной здесь модели  последнего и вопреки
хрестоматийному мнению, результаты этих алгоритмов не
тождественны, причем первый противоречив, а второй~--- нет. Часть
описанных результатов изложена в \cite{bib5}.

Аббревиатуры: КЧ~--- квантовая частица; ВФ~--- волновая функция;
РКЧ~--- релятивистская КЧ; КМ~--- квантовая механика; РКМ~---
релятивистская КМ;  СТО~--- специальная теория относительности;
РКТ~--- релятивистская квантовая теория; КП~---  квантовое поле;
ЭКФ~--- эрмитова квадратичная форма; ВУ~---  волновое уравнение;
ВО~--- волновой оператор.

\pagebreak

\end{document}